\documentclass[aps,amsfonts,prl,nobibnotes,nofootinbib, %
tightenlines, ]{revtex4}
\newcommand{\be}{\begin{eqnarray}}
\newcommand{\ee}{\end{eqnarray}}
\newcommand{\hk}{\hspace{0.1cm}}

\newcommand{\rk}{\right)}
\newcommand{\lk}{\left(}

\newcommand{\il}{\int\limits}

\begin{document}
\tighten

\title{On 't Hooft's loop operator\footnote{supported by DFG - RE 856/5-1}}
\author{H. Reinhardt}
\address{Institut f\"ur Theoretische Physik\\
Universit\"at T\"ubingen\\
Auf der Morgenstelle 14\\
D-72076 T\"ubingen\\
Germany}


\begin{abstract}
An
explicit realization of 't Hooft's loop operator in continuum Yang-Mills theory
is given.
\end{abstract}

\maketitle


\section{Introduction}

In a gauge theory all physical information is contained in (all possible) Wilson
loops
\be
\label{G1}
W [A] (C) = P \exp \lk - \oint\limits_C d x_\mu A_\mu (x) \rk \hk ,
\ee
where $A$ is the algebra valued gauge field\footnote{Strictly speaking it would
be sufficient to consider the trace of $W [A] (C)$. Note also, that we 
use anti-hermitian
generators of the gauge group.}.
This is the basis of the so-called loop space formulation of gauge theory. The
expectation value of a temporal Wilson loop $<W (C) >$ shows an area law in the
confined phase, but screening of the color charges carried by the loop
in the deconfined phase. Thus the temporal Wilson loop can serve
as an order parameter to distinguish the two phases of Yang-Mills theory. 

Another order (or more precisely ``disorder'') 
parameter of Yang-Mills theory was introduced by 't Hooft
and is commonly refered to as 't Hooft loop. This (dis-) 
order parameter  is in a certain sense
dual to the Wilson loop and is defined in the following way:
Consider Yang-Mills theory in the Hamiltonian formulation (which assumes Weyl gauge
$A_0 = 0$), where the spatial components of the gauge field $\vec{A} (\vec{x})$
are the coordinates of the theory. (More precisely in this context
 $\vec{A} (\vec{x})$ has to be
understood as the quantum mechanical ``operator'' 
of the canonical coordinate. In the following we will indicate quantum
mechanical operators by a
hat ``\^{ }''). The operator for a spatial
Wilson loop $\hat{W} (C)$ is given by eq. (\ref{G1}) 
as $\hat{W} (C) = W [\hat{A}] (C)$,
where $C$ is a loop in ordinary three-space $\mathbb{R}^3$. The operator of
't Hooft's loop $\hat{V} (C)$ is then defined by its commutation relation with
the operator of the Wilson loop (\ref{G1}):
\be
\label{G5}
\hat{V} \lk C_1 \rk \hat{W} \lk C_2 \rk = Z^{L \lk C_1, C_2 \rk} \hat{W} \lk C_2
\rk \hat{V} \lk C_1 \rk \hk .
\ee
Here $Z$ denotes an element of the center of the gauge group (which is $Z (N)$
for gauge group $SU (N)$) and 
\be
\label{3}
L \lk C_1, C_2 \rk = \frac{1}{4 \pi} \oint\limits_{C_1} d x_i \oint\limits_{C_2}
d y_j \epsilon_{i j k} \frac{x_k - y_k }{| \vec{x} - \vec{y} |^3}
\hk 
\ee
is the Gaussian linking number between the two spatial loops $C_1, C_2$.

Like the temporal Wilson loop the expectation value of the (spatial) 
't Hooft loop can serve as an
order parameter to distinguish the two phases of Yang-Mills theory. As argued by
't Hooft \cite{1} in the
confined phase the expectation value of the spatial 
't Hooft loop operator $< \hat{V} (C) >$ satisfies a perimeter law (at
zero temperature), while in the deconfined phase it shows an area
law \cite{1}. 
Its behavior is thus dual to the one of the expectation value of a temporal Wilson
loop. 

The 't Hooft loop operator, defined by eq. (\ref{G5}), can be interpreted as
a center vortex creation operator \cite{1}. 
In statistical physics operators, creating 
 topological excitations, like vortices, are referred to
as disorder operators. Their expectation
values, referred to as disorder parameters, are related to the free energy
of the associated topological excitations. 

An explicit realization of a vortex creation operator can be easily given on the
lattice, where center vortices represent co-closed $D - 2$ dimensional
hypersurfaces of plaquettes being equal to a non-trivial center
element\footnote{Strictly speaking the vortex surfaces leave on the dual
lattice.}. The
expectation values of the creation operator of vortices wrapping around the
periodic torus have been studied numerically in ref. \cite{2}. 

Unfortunately 't Hooft 
did not give an explicit representation of its loop operator in
continuum Yang-Mills theory but rather defines this operator by its effect on
the eigenstates of the gauge potential: The effect of ${\hat{V}} (C)$ is a gauge
transformation $\Omega^{[C]}$, which is singular on the curve $C$, and 
if another
curve $C'$ (parametrized by $s \in [0, 1]$) winds through $C$ with $n$ windings
in a certain direction then $\Omega^{[C]}$ receives a phase:
\be
\label{4}
\Omega^{[C]} (s = 1) = e^{i \frac{2 \pi}{N} \cdot n} \Omega^{[C]} (s = 0) \hk .
\ee
This implies that the gauge function $\Omega^{[C]} (x)$ is multivalued but
nevertheless the gauge transformed fields $A^{\Omega^{ [C]}}$ will be single valued.

It is the purpose of the present paper to provide an explicit realization of 't
Hooft's loop operator in continuum Yang-Mills theory. In the following we will
first construct an operator that
generates the singular gauge transformations defined by eq. (\ref{4}). Second we
will show that this operator, when acting on physical (i.e. gauge invariant)
states, creates a center vortex. Third we will construct a center vortex
creation operator and explicitly show that it satisfies indeed the defining
equation (\ref{G5}) for 't Hooft's loop operator.

\section{Generator of singular gauge transformation}

In the following we explicitly construct the singular gauge transformation
defined by eq. (\ref{4}). 
A good candidate for this
gauge function is
\be
\label{5}
\Omega (\Sigma , x) = e^{- E \omega (\Sigma, x)} \hk .
\ee
Here 
$E = E_a
T_a$ denotes a co-weight vector in the Lie algebra, which when exponentiated
produces a non-trivial center element
\be
\label{6}
e^{- E} = Z \hk .
\ee
Furthermore $ \omega (\Sigma, x)$ is the solid angle subtended by the 
loop $\partial
\Sigma$ as seen from the point $x$, which can be expressed as 
\be
\label{Ga4}
\omega (\Sigma, x) = \il_\Sigma d^2 \tilde{\sigma}_i \partial^x_i D \lk x -
\bar{x} (\sigma) \rk \hk ,
\ee
where $D (x)$ denotes the Green's function of the 3-dimensional Laplacian,
satisfying $- \partial^2 D (x) = \delta^3 (x)$.
(We have
normalized $\omega (\Sigma, x)$ so that the total solid angle of a 2-sphere
$S^2$ is unity.) 
A deformation of $\Sigma$ keeping its boundary $C
= \partial \Sigma$ fixed, leaves $\omega (\Sigma , x)$ 
invariant, unless $x$ crosses
$\Sigma$. When $x$ crosses the surface $\Sigma$ the solid angle $\omega (\Sigma,
x)$ jumps by $\pm 1$ and accordingly the gauge function (\ref{5}) changes  by
a center element, eq. (\ref{6}).

However, as defined by  (\ref{5}), $\Omega
 (\Sigma , x)$ is a single valued function.
To obtain 't Hooft's multivalued
gauge function $\Omega^{[C]}$ (\ref{4}) from $\Omega (\Sigma , x)$ (\ref{5}), 
we have to ignore the discontinuity (jump) of $\omega
(\Sigma, x)$, when $x$ crosses $\Sigma$ and instead use the smooth (but
multivalued) continuation of  $\omega (\Sigma, x) \to \omega^{[\partial \Sigma =
C]}$ (with the jump of $x$ at $\Sigma$ ignored), which then depends only on
$\partial \Sigma = C$. Let us now explicitly construct the generator of the
gauge transformation (\ref{4}).

The operator, which generates a 
gauge transformations $\Omega 
= e^{- \Theta} , \Theta = \Theta^a T_a$ of the field
operators is given by
\be
\label{Ga1}
{\cal U} (\Theta) = exp \left[ i \int d^3 x \lk \hat{D}^{a b}_i (x) \Theta^b \rk
\Pi^a_i (x) \right] \hk ,
\ee
where $\hat{D}_i = \partial_i + \hat{A}_i$ denotes the covariant derivative with
$\hat{A}^{a b}_i = f^{a c b} A^c_i$ being the gauge field in the adjoint
representation and $f^{a b c}$ denotes the structure
constants of the gauge group. 
Furthermore
\be
\label{G12}
\Pi^a_i (x) = \frac{1}{i} \frac{\delta}{\delta A^a_i (x)} \hk .
\ee
is the ``momentum operator'' in Yang-Mills theory, which coincides with the
operator of the color electric field and satisfies the (equal time) 
canonical commutation
relation
\be
\label{ZZ}
\left[ A^a_k \lk \vec{x} \rk \hk , \hk \Pi^b_l \lk \vec{y} \rk \right] = i
\delta^{a b} \delta_{ k l} \delta^3 \lk \vec{x} - \vec{y} \rk \hk .
\ee
Any operator in quantum Hilbert space is gauge
transformed by the action of the operator (\ref{Ga1}). For example, for the
coordinate operator we have, by using (\ref{ZZ}),
\be
\label{Ga2}
{\cal U} (\Theta) A {\cal U}^\dagger (\Theta) = 
\Omega A \Omega^\dagger + \Omega \partial \Omega^\dagger = A^\Omega \hk
 .
 \ee

By putting $\Theta = E \omega (\Sigma, x)$, we obtain the operator which
generates the gauge transformation (\ref{5})
\be
\label{X1a}
{\cal U} ( E \omega) = exp \left[ i \int d^3 x E^a \lk \partial_i \omega \rk \Pi^a_i +
\omega \hat{A}^{a b}_i E^b \Pi^a_i \right] \hk .
\ee
This operator 
leaves physical states invariant. To see this perform a partial integration in
the exponent of eq. (\ref{X1a})
\be
\label{Ga6}
{\cal U} (E \omega) = exp \left[ i \int d^3 x \lk \partial_i \lk \lk E \omega \rk^a
\Pi^a_i (x) \rk - \omega E^a \hat{D}^{a b}_i \Pi^b_i (x) \rk \right] \hk .
\ee
The first term in the exponent can be converted into a surface integral, which
vanishes as the solid angle $\omega (\Sigma, x)$ vanishes for $\vec{x}^2 \to
\infty$\footnote{In fact, this term can be shown to vanish for all topologically
trivial gauge transformation with zero winding number $n [\Omega] \in \Pi_3 \lk S_3
\rk$}. The second term in the exponent of eq. (\ref{Ga6}) vanishes, when acting
on gauge invariant states by Gauss' law $\hat{D} \Pi | \Psi > = 0$. Thus, when
taking the operator ${\cal U} (E \omega)$ literally it has no effect on physical
states. However, this operator becomes non-trivial, when one neglects the
singular part of $\partial_i \omega$, which arises from the jump of $\omega
(\Sigma, x)$ by $\pm 1$, when $x$ crosses the surface $\Sigma$.
Then this operator
generates precisely the desired multivalued gauge transformation (\ref{4}), 
which defines
the effect of 't Hooft's loop operator. Thus eq. (\ref{X1a}), ignoring the
singular part of $\partial_i \omega$ should give an explicit realization of 't
Hooft's loop operator. In fact such an operator (\ref{X1a}) was considered in
ref.\cite{4} for planar loops as an explicit realization of the 't Hooft loop
operator, but it was not shown that this operator satisfies the defining eq.
(\ref{G5}).
In this paper we will explicitly show, that the so defined operator (\ref{X1a})
(again with the singular piece of $\partial_i \omega$ omitted) is gauge
equivalent to a center vortex creation operator and that the
latter will satisfy 't Hooft's loop algebra (\ref{G5}).

\section{Center vortex creation operator}

Let us now explicitly work out the effect of ignoring the singular part of
$\partial \omega$. One can show that \cite{3}
\be
\label{XX}
\Omega (\Sigma, x) \partial \Omega^\dagger (\Sigma, x) = E \partial \omega 
(\Sigma, x) 
= a ( \partial \Sigma, x) - {\cal
A} (\Sigma, x) \hk ,
\ee
where
\be
\label{Ga8}
{\cal A}_i (\Sigma, x) = 
E \int_\Sigma d^2 \tilde{\sigma}_i \delta^3 \lk x - \bar{x}
(\sigma) \rk \hk ,
\ee
\be
\label{Ga9}
a_i (\partial \Sigma, x) = 
E \int_{\partial \Sigma} d  \bar{x}_k \epsilon_{k i j}
\partial^{\bar{x}}_j D \lk x - \bar{x} \rk  \hk .
\ee
Here 
$\bar{x}_i (\sigma)$  denotes a parameterization of the 2-dimensional 
open
surface $\Sigma$.
Furthermore 
$a (\partial \Sigma, x)$ and ${\cal A} (\Sigma, x)$ represent the smooth
and the singular part of $E \partial \omega$, respectively. The singular part
${\cal A} (\Sigma, x)$ arises precisely from the jump of $\omega (\Sigma, x)$,
when $x$ crosses $\Sigma$. From the physical point of view $a (\partial \Sigma,
x)$ and  ${\cal A} (\Sigma, x)$ are different gauge representations of one and 
the same
center vortex located at $C = \partial \Sigma$.
An ideal center vortex field ${\cal A} (C_1)$, whose flux is located at a loop
$C_1$ is defined by its property that it
produces a non-trivial center element for Wilson loops $C_2$ non-trivially
(modulo N) linked to the loop $C_1$, i.e.
\be
\label{G16}
W \left[ {\cal A} \lk C_1 \rk \right] \lk C_2 \rk = Z^{L \lk C_1, C_2 \rk} \hk ,
\ee
where $L \lk C_1, C_2 \rk$ is the Gaussian linking number (\ref{3}). Indeed the
two potentials (\ref{Ga8}), (\ref{Ga9}) satisfy
\be
\oint\limits_C d x_i {\cal A}_i \lk \Sigma, x  \rk & = & E I (C, \Sigma)
\nonumber\\
\oint\limits_C d x_i a_i \lk \partial \Sigma, x  \rk & = & E L (C, \partial 
\Sigma) \hk ,
\ee
where $I (C, \Sigma)$ is the intersection number between the loop $C$ and the
open surface $\Sigma$. Since $I (C, \Sigma) = L (C, \partial \Sigma)$ both
potentials satisfy eq. (\ref{G16}).
Note, that ${\cal A} (\Sigma, x)$ represents the continuum analog of the ideal
center vortices \cite{3} arising on the lattice after center projection \cite{RX}, 
which in D = 3 gives rise to open surfaces $\Sigma$ of links being
equal to a non-trivial center element $Z \in Z (N)$. 
The boundaries $C = \partial \Sigma$ of
these surfaces define closed loops of center flux on the dual lattice. $a
(\partial \Sigma, x)$ (\ref{Ga9}) 
represents the same center vortex as ${\cal A} (\Sigma,
x)$ (\ref{Ga8}), 
but depends only on the vortex loop $C = \partial \Sigma$, and is
transversal $\vec{\partial} \vec{a} (\partial \Sigma, x) = 0$.

Neglecting the singular part of $\partial \omega (\Sigma, x)$ in the exponent of
eq.(\ref{X1a}), i.e. neglecting ${\cal A} (\Sigma, x)$ in eq. (\ref{XX}),
we obtain the operator 
\be
\label{Ga10}
{\bar{ \cal{U}}} (E \omega) = e^{i \int d^3 x \lk a^a_i \Pi^a_i  - \omega E^a
\hat{A}^{a b} \Pi^b_i \rk} \hk ,
\ee
which precisely generates the singular gauge transformation (\ref{4}), i.e.
${\bar{\cal U}} (E \omega) A {\bar{\cal U}}^\dagger (E \omega) = 
A^{\Omega^{[\partial \Sigma]}}$. For later use we rewrite this
operator by
adding a ``zero'' in the exponent and using eq. (\ref{XX}) 
\be
\label{Ga11}
\bar{\cal U} (E \omega) & = & exp \left[ i \int d^3 x \left[ \lk a^a_i - {\cal A}^a_i
\rk \Pi^a_i 
 - \omega E^a \hat{A}^{a b}_i \Pi^b_i + {\cal A}^a_i \Pi^a_i
(x) \right] \right] \nonumber\\
& = & exp \left[ i \int d^3 x \left[ {\cal A}^a_i (x) \Pi^a_i (x) + \lk
\hat{\cal D}_i \omega E \rk^a \Pi^a_i \right] \right]  \hk .
\ee
Now observe that the two operators in the exponent commute
\be
 \left[ {\cal A}^a_k \Pi^a_k , \lk \hat{D} \omega \rk^b \Pi^b \right]
 =  {\cal A}^a_k \left[ \Pi^a_k , \lk \hat{A} \omega \rk^b \right] \Pi^b
 =  {\cal A}^a_k f^{b a c} E^c \omega \Pi^b_k = 0 \hk .
\ee
The last relation follows since ${\cal A}^a_k \sim E^a$ (see eq. (\ref{Ga8}). 
We can
therefore rewrite the operator (\ref{Ga11}) as
\be
\bar{\cal U} ( E \omega) = exp \left[ i \int d^3 x {\cal A}^a_i (\Sigma, x)
\Pi^a_i (x) \right] {\cal U} (E \omega) \hk ,
\ee
where ${\cal U} (E \omega)$ is the generator (\ref{Ga1}) of the (single valued)
gauge transformation $\Omega (\Sigma,x)$ (\ref{5}). 
When acting on physical states $| \Psi >$ this
operator becomes unity  (${\cal U} (E \omega) | \Psi > = | \Psi >$) 
and the operator ${\bar{\cal U}} (E
\omega)$ can be replaced by
\be
\label{21}
{\hat V} (\Sigma) = \exp \left[ i \int d^3 x {\cal A}^a_i \lk \Sigma , x \rk
\Pi^a_i (x) \right] \hk .
\ee
In the next section we will show, that this operator indeed satisfies the
defining equation (\ref{G5}) for 't Hooft's loop operator. Before giving that
proof
let us show that this operator generates a center vortex located at $\partial
\Sigma$.

A vortex creation operator should add a center vortex field to any given gauge
field $\vec{A} (x)$. Let us recall, that in the canonical formulation of continuum
Yang-Mills theory the gauge field $\vec{A} (x)$ represents the quantum
mechanical coordinate. Let $| A >$ denote an eigenstate of the quantum
mechanical operator $\hat{A}$ of the Yang-Mills
coordinate with ``eigenvalue'' 
$\vec{A} (x)$ i.e.
\be
\hat{\vec{A}} (x) | A > = \vec{A} (x) | A > \hk .
\ee
A vortex creation operator should then satisfy
\be
\label{G8}
\hat{V} (C) | A > = | A + {\cal A} (C) > \hk ,
\ee
where ${\cal A} (C)$ denotes the gauge potential of a center vortex located at the
loop $C$. In quantum mechanics the operator, which shifts the coordinate, say
$x$, by an amount ${x_0}$
\be
T \lk x_0 \rk  | x > = | x + x_0 > \hk 
\ee
is given by
\be
\label{G10}
T \lk x_0 \rk = e^{i x_o \hat{p} } \hk ,
\ee
where $\hat{p} = \frac{1}{i} \frac{d}{d x}$ is the usual momentum operator. 
The operator (\ref{G10}) obviously generates a translation in coordinate space.
Analogously the operator (\ref{21}) 
creates a center vortex field ${\cal A} (\Sigma)$.
Indeed from the commutation relation (\ref{XX}) follows 
\be
\label{G14}
\hat{V} (C) \hat{A} \hat{V} (C)^\dagger = \hat{A} + {\cal A} (C) \hk 
\ee
and by Taylor expansion one finds for any functional of the gauge potential
\be
\label{G15}
\hat{V} (C) f \lk \hat{A} \rk \hat{V} (C)^\dagger = f \lk \hat{A} + {\cal A} (C)
\rk \hk,
\ee
which proves eq. (\ref{G8}). Eq. (\ref{21}) 
represents the continuum version of the 
center vortex analog of the monopole creation operator
introduced by DiGiacomo at al. on the lattice \cite{5} .

Inserting the explicit representation (\ref{Ga8}) into eq. (\ref{21}) 't
Hooft's loop operator becomes
\be
\label{G16a}
\hat{V} (C) = e^{i \il_{\Sigma (C)} d^2 \sigma_k E^a \Pi^a_k \lk \bar{x}
(\sigma) \rk}  \hk .
\ee
Since $\Pi^a_i (x)$ represents the operator of the electric field it is seen
that the 't Hooft loop operator $\hat{V} (C)$ measures the electric flux through
the loop $C$. In this sense the spatial 't Hooft loop is indeed dual to the
spatial Wilson
loop, which measures the magnetic flux.


\section{Proving 't Hooft's loop algebra}

Let us now show that the vortex creation operator defined by eqs. (\ref{21}),
(\ref{G12})
and (\ref{Ga8}), indeed, satisfies the defining eq. (\ref{G5}) for the 't Hooft
loop operator. 

>From eq. (\ref{G15}) follows
\be
\label{G19}
\hat{V} \lk C_1 \rk W [\hat{A}]  \lk C_2 \rk \hat{V} \lk C_1 \rk^\dagger = 
W \left[ \hat{A}
+ {\cal A}  \lk C_1 \rk \right] \lk C_2 \rk \hk .
\ee
Note, that $\hat{V} (C)$ is an operator in quantum Hilbert space but not in
color space (it does not contain the generators of the gauge group). Hence, this
operator does not interfer with the path ordering in the Wilson loop. However,
in the resulting Wilson loop $W \left[ \hat{A}
+ {\cal A}  \lk C_1 \rk \right] \lk C_2 \rk$ appearing on the 
right-hand side of 
eq.(\ref{G19}) the two gauge potentials $\hat{A}$ 
and ${\cal A} \lk C_1 \rk$ are both Lie algebra
valued gauge fields, which in general do not commute. 

To simplify the expression
for the Wilson loop $W \left[ A
+ {\cal A}  \lk C_1 \rk \right] \lk C_2 \rk$ let us introduce a parameterization 
of the loop
$C_2 \hk : \hk \bar{x} (s) \hk , \hk s \in [0, 1]$. Then the Wilson loop
\be
\label{G17a}
W [A] \lk C_2 \rk = P e^{- \il^1_0 d s \dot{\vec{x}} (s) \vec{A} (x (s))} 
\ee
can be interpreted as a ``time evolution operator'' in color space. 
Adopting for this evolution
operator the familiar ``interaction picture'', interpreting $\dot{\vec{x}} 
(s)
\vec{\cal A} \lk C_1, \vec{x} (s) \rk$  and  
$\dot{\vec{x}} (s)
\vec{A} (x (s))$ as 
``free'' and ``interaction'' parts, respectively,  of the ``Hamiltonian'' 
we find
\be
\label{G21} 
 W \left[ A + {\cal A} \lk C_1 \rk \right] \lk C_2 \rk 
 =  W \left[ {\cal A} \lk
C_1 \rk \right] \lk C_2 \rk W \left[ A^I \lk C_1 \rk \right] \lk C_2 \rk \hk ,
\ee
where
\be
\label{G22}
 A^I \lk C_1 \hk ; \hk x (s) \rk
  =  U^\dagger \lk C_1 \hk , \hk s \rk A \lk x
(s) \rk U \lk C_1 , s \rk \hk ,
\ee
\be
\label{35}
 U \lk C_1 , s \rk 
 =  U \left[ {\cal A} (C_1) \right] (s)
  =  P 
\exp \lk - \il^s_0 d s' \dot{\vec{x}} (s') \vec{\cal A}
\lk C_1 , x (s') \rk  \rk 
\ee
is the interaction representation of the color matrix $A (x) = A^a (x) T_a$. 
Note, that in the
``free'' time evolution operator $U \lk C_1, s \rk$ defined by eq. (\ref{35})
the path ordering is irrevelant, since the vortex gauge potential $\vec{\cal A}
= {\cal A}^a T_a$ defined by eq.
(\ref{Ga8}) lives in the Cartan subalgebra. (For later use we have, however,
kept the path ordering operator.) With the explicit expression for the
vortex gauge potential, eq. (\ref{Ga8}), we find for the ``free'' time evolution
operator (\ref{35}).
\be
\label{36}
U \lk C_1 ;  s  \rk  = e^{- E I \lk L_{x (0) \rightarrow x (s)} , \Sigma
\lk C_1 \rk \rk} \hk ,
\ee
where
\be
\label{G24}
I \lk L_{x (0) \longrightarrow x (s)} , \Sigma \rk = \il^{x (s)}_{x (0)} d
\vec{x} \il_\Sigma d \vec{\sigma} \delta^3 \lk x - \bar{x} (\sigma) \rk \hk 
\ee
is the intersection number between the open path segment $L_{x (0)
\longrightarrow x (s)}$ of the loop $C_2$ and the open surface $\Sigma \lk C_1
\rk$ bounded by $C_1$. Note, the intersection number, eq. (\ref{G24}), depends on
the particularly chosen surface $\Sigma \lk C_1
\rk$. Different choices of $\Sigma \lk C_1
\rk$ keeping the boundary $C_1 = \partial \Sigma \lk C_1 \rk$ fixed corresponds
to different choices of gauges related by a center gauge transformations, see
ref. \cite{3}. However, for any surface 
$\Sigma \lk C_1
\rk$ (i.e. center gauge) choosen the intersection number $I \lk L, \Sigma \lk
C_1 \rk \rk$ is an integer, and accordingly the evolution operator eq.
(\ref{35}) is equal to a center element (\ref{6})
\be
\label{22a}
U \lk C_1 , x (s) \rk = Z^{I \lk L_{x (0) \to x (s)} , \Sigma \lk
C_1 \rk \rk} 
\ee
and, hence, commutes with any gauge potential. This implies, that the
``interaction representation'' $A^I \lk C_1, x \rk$, eq. (\ref{G22}), coincides
with the gauge potential itself, that is $A^I  \lk C_1, x \rk = A (x)$ and eq.
(\ref{G21}) simplifies to
\be
W \left[ A + {\cal A} \lk C_1 \rk \right] \lk C_2 \rk = W \left[ {\cal A} \lk C_1 \rk
\right] \lk C_2 \rk W \left[ A \right] \lk C_2 \rk \hk .
\ee
Using this relation and, furthermore, eq. (\ref{G16}) one finds from eq.
(\ref{G19}) immediately the desired result (\ref{G5}).

A comment is here in order:
Strictly speaking 't Hooft's loop operator is defined by eq. (\ref{G5}) with
$\hat{W} (C_2)$ replaced by its trace, $tr \hat{W} (C_2)$. However, taking the
trace of eq. (\ref{G5}) and taking into account that $\hat{V} (C)$,
(\ref{21}) is a unit matrix in color space precisely replaces $\hat{W} (C)$ 
by $tr \hat{W} (C)$.

\section{Gauge invariance}

Under (non-singular) gauge transformation $\Omega: 
A \longrightarrow A^\Omega = \Omega A \Omega^\dagger + \Omega \partial
\Omega^\dagger $
the operator of the electric field transforms homogeneously
\be
\Pi = \Pi^a T_a \longrightarrow \Pi^\Omega = \Omega \Pi \Omega^\dagger \hk .
\ee
Thus a gauge transformation de facto replaces the center vortex field ${\cal A} =
{\cal A}_a T_a$ in the 't Hooft loop operator $\hat{V} (C)$ (\ref{21}) by
the color rotated one $
\Omega^\dagger {\cal A} \Omega \hk .
$
Thus in order to prove that the
result of the action of the 't Hooft operator (\ref{21}) 
is independent of the chosen gauge\footnote{From eq. (\ref{G8}) and (\ref{G14})
it is seen that 't Hooft's loop operator shifts the canonical coordinate
$\vec{A}$ by the center vortex field. Requiring the shifted field variable
$\vec{A'} = \vec{A} + \vec{\cal A} [C]$ to transform under gauge transformations
in the same way as the unshifted one, $\vec{A}$, (see eq. (\ref{Ga2})) 
implies to
demand the ``background'' field ${\cal A} (C)$ to transform homogeneously,
${\cal A} (C) \longrightarrow \Omega {\cal A} (C) \Omega^\dagger$. Then the 't
Hooft operator would be manifestly gauge-invariant.}
we have to show that the 't Hooft algebra (\ref{G5}) remains
 valid when the center
vortex field ${\cal A} (C, x)$ is replaced by $\Omega^\dagger {\cal A} (C, x)
\Omega$. For this we have to show that ${\cal A} (C)$ and 
$\Omega^\dagger {\cal A} (C) \Omega$ 
give rise to
the same evolution operator (\ref{35})
\be
\label{G28}
U \left[ \Omega^\dagger {\cal A} (C) \Omega \right] (s) = U \left[ {\cal A} (C)
\right] (s) \hk .
\ee
Since for $s = 1$ the evolution operator $U \left[ {\cal A} (C) \right] (s)$
(\ref{G22}) becomes the Wilson loop $W \left[ {\cal A} (C) \right] (C_2)$,
eq. (\ref{G1}), from eq. (\ref{G28}) follows that $\Omega^\dagger {\cal A} (C,
x) \Omega$ 
and ${\cal A} (C, x)$ represent the same center vortex flux, i.e.
\be
W  \left[ \Omega^\dagger {\cal A} (C_1) \Omega \right] \lk C_2 \rk = W
\left[{\cal A} \lk C_1 \rk \right] \lk C_2 \rk = Z^{L \lk C_1, C_2 \rk} \hk .
\ee
To prove eq. (\ref{G28}) 
 we write $\Omega^\dagger {\cal A} \Omega$ as gauge transformed of 
${\cal
A}^\dagger + \Omega \partial \Omega^\dagger$ i.e.
\be
\Omega^\dagger {\cal A} \Omega = \lk {\cal A} + \Omega \partial
\Omega^\dagger\rk^{\Omega^\dagger} \hk .
\ee
>From the definition (\ref{35}) follows
\be
\label{X1}
U \left[ A^{\Omega} \right] (s) = \Omega (s) U \left[ A \right]
(s) \Omega^\dagger (0) \hk .
\ee
Hence we have
\be
\label{X3}
U \left[ \Omega^\dagger {\cal A} \Omega \right] (s) = U \left[ \lk 
{\cal A} + \Omega
\partial \Omega^\dagger \rk^{\Omega^\dagger} \right] = \Omega^\dagger (s) U
\left[ {\cal A} + \Omega \partial
\Omega^\dagger \right] (s) \Omega (0) \hk .
\ee
For the evolution operator $U \left[ {\cal A} (C) + \Omega \partial
\Omega^\dagger \right]$ we use again the ``interaction representation'' with
${\cal A} (C)$ and $B =  \Omega \partial
\Omega^\dagger$ being the ``free'' and ``interaction'', respectively,
parts
\be
\label{X2}
U \left[ {\cal A} (C) 
\right] (s) = U \left[ {\cal A} (C) \right] (s) U \left[ B^I
(C) \right] (s) \hk ,
\ee
where the interaction representation $B^I (C)$ is defined by eqs. (\ref{G22}),
(\ref{35}).
Using the fact that $U \left[ {\cal A} (C) \right] (s)$ represents a center
element, see eq. (\ref{22a}), we find that $B^I (C) = B = \Omega \partial
\Omega^\dagger$, so that eq. (\ref{X2}) becomes
\be
\label{X4}
U \left[ {\cal A} (C) + \Omega \partial
\Omega^\dagger \right] (s) = U \left[ {\cal A} (C) \right] (s) U \left[ 
\Omega \partial
\Omega^\dagger \right] (s) \hk .
\ee
Using eq. (\ref{X1}) we have
\be
U \left[ \Omega \partial
\Omega^\dagger \right] (s) = \Omega (s) U [0] (s)
\Omega^\dagger (0) \hk ,
\ee
with $U [0] (s) = 1$, so that we find from (\ref{X3}) and (\ref{X4})
\be
U \left[ \Omega^\dagger {\cal A} \Omega \right] (s) 
& = & \Omega^\dagger (s) \lk U
\left[ {\cal A} (C) \right] (s) \Omega (s) \Omega^\dagger (0) \rk 
\Omega (0)
\nonumber\\
& = & U \left[ {\cal A} (C) \right] (s) \hk ,
\ee
where we have again used that $U \left[ {\cal A} (C) \right] (s)$ represents a
center element (see eq. (\ref{36}). 
This proves that the vortex generation operator $\hat{V} (C) $
(\ref{G8}) does not change when the center vortex field ${\cal A} (C)$ is
replaced by the color rotated 
one $\Omega^\dagger {\cal A} (C) \Omega$, which proves
the gauge invariance of the action of $V (C)$.

\section{Acknowledgments} 

Discussions with T.~Tok and L.~v.~Smekal are greatly acknowledged.


\end{document}